\documentclass[letter]{aa}    

\usepackage{graphicx}
\usepackage{txfonts}
\usepackage{lipsum}
\usepackage{subcaption}         
\usepackage{lscape}             
\usepackage{placeins}           
\usepackage{hyperref}

\hypersetup{colorlinks=true,
urlcolor=blue, 
linkcolor=blue,
citecolor=blue}

\begin{document}

   \title{Mare versus highland lunar impact flash light curve dichotomy}

   \subtitle{}

   \author{D. Athanasopoulos\inst{1} \corrauth{dimathanaso@noa.gr}
        \and A. Liakos\inst{1} \email{alliakos@noa.gr}
        \and A. Z. Bonanos\inst{1} \email{bonanos@noa.gr}
        \and D. Koschny\inst{2} \email{detlef.koschny@tum.de}
        \and \\ O. Sykioti\inst{1} \email{sykioti@noa.gr}
        \and M. Devog\`ele\inst{3} \email{Maxime.Devogele@ext.esa.int}
        \and J. L. Cano\inst{4} \email{Juan-Luis.Cano@esa.int}
        \and R.~Moissl\inst{4} \email{richard.moissl@esa.int}
        }

   \institute{IAASARS, National Observatory of Athens, 15236 Penteli, Greece
   \and Technical University of Munich, Lunar and Planetary Exploration, Lise-Meitner-Str. 9, 85521 Ottobrunn, Germany
   \and ESA NEO Coordination Centre, European Space Agency, Largo Galileo Galilei, 1, 00044, Frascati, RM, Italy
   \and ESA ESOC, Planetary Defence Office, Robert-Bosch-Strasse 5, 64293 Darmstadt, Germany}

   \date{Received June 4, 2026}

\abstract{We performed a comprehensive analysis of lunar impact flash (LIF) light curve shapes and their dependence on the lunar terrain, using the large sample of LIFs detected by NELIOTA over the last 9 years. We classified 124 multi-frame light curves into mare, highland, and `border' regions. Subsequently, we derived analytical expressions for single-size and dual-size ejecta cooling models, which we fitted to the observational data to estimate their physical properties. While impacts on the two terrains yield similar peak magnitude distributions, their decay behaviour differs significantly; highland LIFs exhibit a shallower and longer-lasting decay ($\tau = 0.11_{-0.05}^{+0.08}$~s) compared to mare flashes ($\tau = 0.038_{-0.018}^{+0.032}$~s). The dual-size model suggests this extended duration is primarily driven by the coarse particles of the ejecta. The profile and duration of the LIF light curves represent the initial stages of the impact cratering process. The observed dichotomy between highland and mare LIFs demonstrates that the initial stages of the impact cratering process are fundamentally dependent on lunar lithology.}


   \maketitle

   \date{Received May 15, 2026}

\nolinenumbers

\section{Introduction}

Hypervelocity impacts are among the most energetic processes in the Solar System, yet the physics of the initial cratering process remains inaccessible to laboratory replication. The Moon, where meteoroid strikes occur at speeds far exceeding experimental limits, serves as a premier natural laboratory. On airless bodies, impacts of meteoroids produce luminous transient events in the visible and near-infrared during the initial stages of the cratering process. Photons are produced by the partial conversion of the impactor's kinetic energy into thermal radiation, which is governed by the critical parameter known as luminous efficiency \citep{bouley2012}. While lunar surface lithology is known to shape late-stage crater morphology, its influence on the initial stages of impact and the resulting lunar impact flashes (LIFs) has remained elusive. 

While unverified reports of such phenomena have existed for centuries \citep[][and references therein]{kolovos1988}, the modern era of lunar impact monitoring began with the 1999 Leonid meteor stream \citep{ortiz2000, artemeva2001, yanagisawa2002, cudnik2003a}. Since then, many observing campaigns have been performed \citep[e.g.][]{ortiz2002,ortiz2006, ortiz2015, cudnik2003b, yanagisawa2006, yanagisawa2021, yanagisawa2025, suggs2008,suggs2014, aitmoulalabri2015, bonanos2018, madiedo2018, madiedo2019, liakos2020, liakos2024, sheward2024}, transforming the Moon into a natural laboratory for hypervelocity impact physics. 

Studies of impact craters on the Earth and the Moon have demonstrated that the target composition, specifically the lithology, plays a significant role in the impact cratering process \citep[see][and references therein]{osinski2019}. However, no equivalent study has examined its potential effect on LIF properties. The lunar surface is morphologically heterogeneous: the high-albedo regions (i.e. highlands) represent the primary lunar crust, whereas the low-albedo maria are basaltic plains formed by subsequent volcanism and exhibit a vastly different mineralogy \citep[e.g.][]{borg2023}. In this study, we leveraged the large LIF sample produced by the NELIOTA programme \citep{xilouris2018, bonanos2018, liakos2020, liakos2024} to examine the dependence of the LIF light curve behaviour on the lunar terrain.

  \begin{figure*}[!hbtp]
  \centering
    \includegraphics[width=0.70\textwidth]{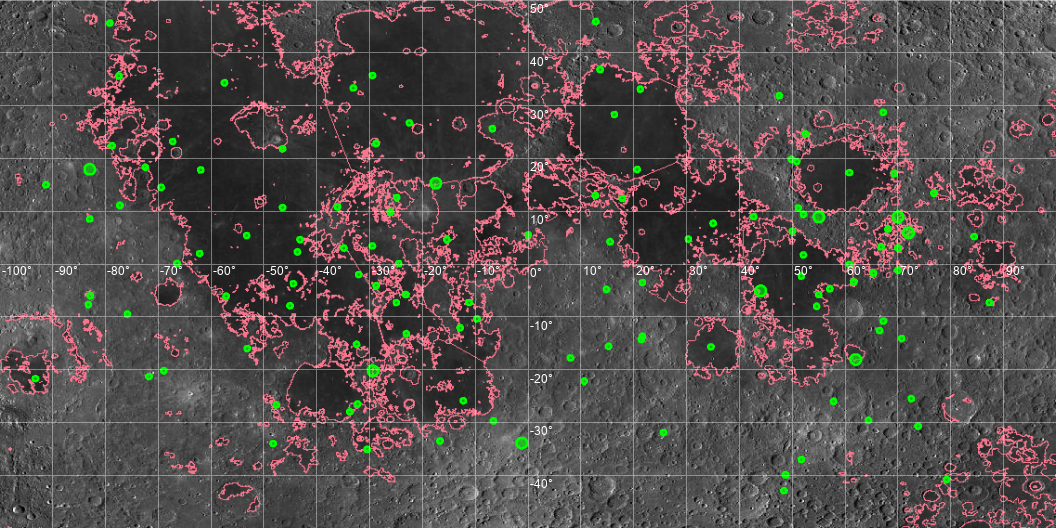}
      \caption{NELIOTA multi-frame LIF sample (green ellipses, scaled according to their localisation accuracy) superimposed on a lunar map of the near side. Mare boundaries are highlighted by solid magenta lines.}
      \label{fig:LunarMap}
  \end{figure*}

\section{Multi-frame lunar impact flashes}

\subsection{Datasets} \label{subsec:datasets}

To date, NELIOTA\footnote{\url{neliota.astro.noa.gr}} has produced the largest publicly available LIF dataset, consisting of 329 reported LIFs \citep[][and herein]{liakos2020, liakos2024}. At least 50\% of the total sample comprise at least two frames in the $I$ filter; these are referred to as `multi-frame' events and have a minimum duration of 66~ms; events exceeding 100~ms are hereafter termed `long' events. The majority of LIFs ($\sim$75\%) have not been associated with known meteoroid streams and are called `sporadics'; they are expected to be chondrites, similar to most common meteorites \citep{broz2024}.

We included the 109 archival NELIOTA multi-frame LIFs documented in \cite{liakos2020, liakos2024} and classified as sporadic; these provide a robust sample with a randomised impact velocity, averaging 24~km~s$^{-1}$ \citep{suggs2014}, thereby isolating the influence of the target terrain on the resulting flash. To this dataset we added 15 of the 34 LIF detections recorded between August 2025 and February 2026 (Table~\ref{tab:NewLIFs}), those that were identified as sporadic multi-frame LIFs (Table~\ref{tab:MultiframeLIFs}). In total, our sample consists of 124 multi-frame LIFs, of which 97 have been validated (see Table~\ref{tab:RegionStats}). Figure~\ref{fig:LunarMap} presents a visualisation of the spatial distribution of the LIF sample on a lunar map mosaic generated by \cite{wagner2015}.

\subsection{Terrain classification of the LIF locations}

\begin{figure}[ht!]
\centering
\includegraphics[width=\hsize]{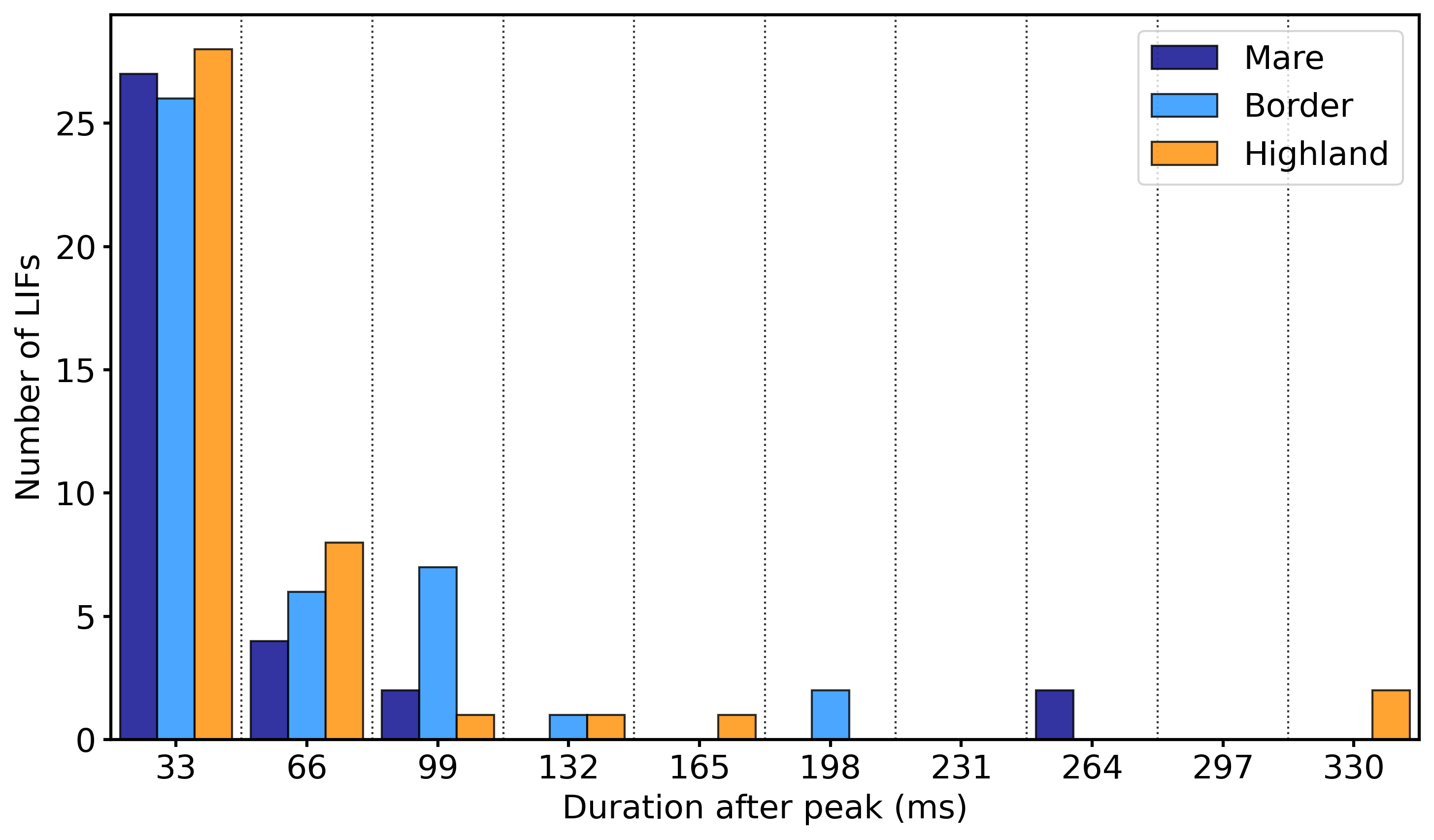}
\caption{Distribution of multi-frame sporadic LIFs vs their duration, grouped by terrain type.}
\label{fig:LIFhist}
\end{figure}

We adopted the grouping scheme laid out in \cite{osinski2019} to classify the LIF sample according to the terrain type of their detected locations: mare, highland, and `border'. For this classification, the mare boundaries defined by \cite{nelson2014} were used (see Fig.~\ref{fig:LunarMap}); a mare or highland event is defined as one where the LIF localisation area, including its associated uncertainty, overlaps a mare or highland region by at least 95\%, respectively. Conversely, `border' LIFs are those that occur near the interface between mare and highland terrains, where the localisation area overlaps the mare boundaries by between 5\% and 95\% (Table~\ref{tab:LIFregions}). Figure~\ref{fig:LIFhist} presents the distribution of the LIF durations grouped by terrain type.

\subsection{Photometry and light curve normalisation}

We performed aperture photometry for the newly detected LIFs following the methodology described by \cite{liakos2020}, and we applied a two-step normalisation process. The first step consisted of a temporal shift, wherein we aligned all peak magnitudes to a relative temporal origin designated as `Frame 1' (at $t_0 = 0$~s). Measurements prior to the peak were matched to `Frame 0'. The second step involved the normalisation of the peak amplitudes; we did this by setting the peak relative magnitude value to zero, which intrinsically corresponds to a maximum normalised flux of 1. This resulting dimensionless representation is paramount for systematically evaluating the characteristic cooling rates, durations, and decay behaviours of sporadic LIFs detected on different lunar terrain types.
\begin{figure*}[ht!]
   \centering
    \includegraphics[width=0.44\textwidth]{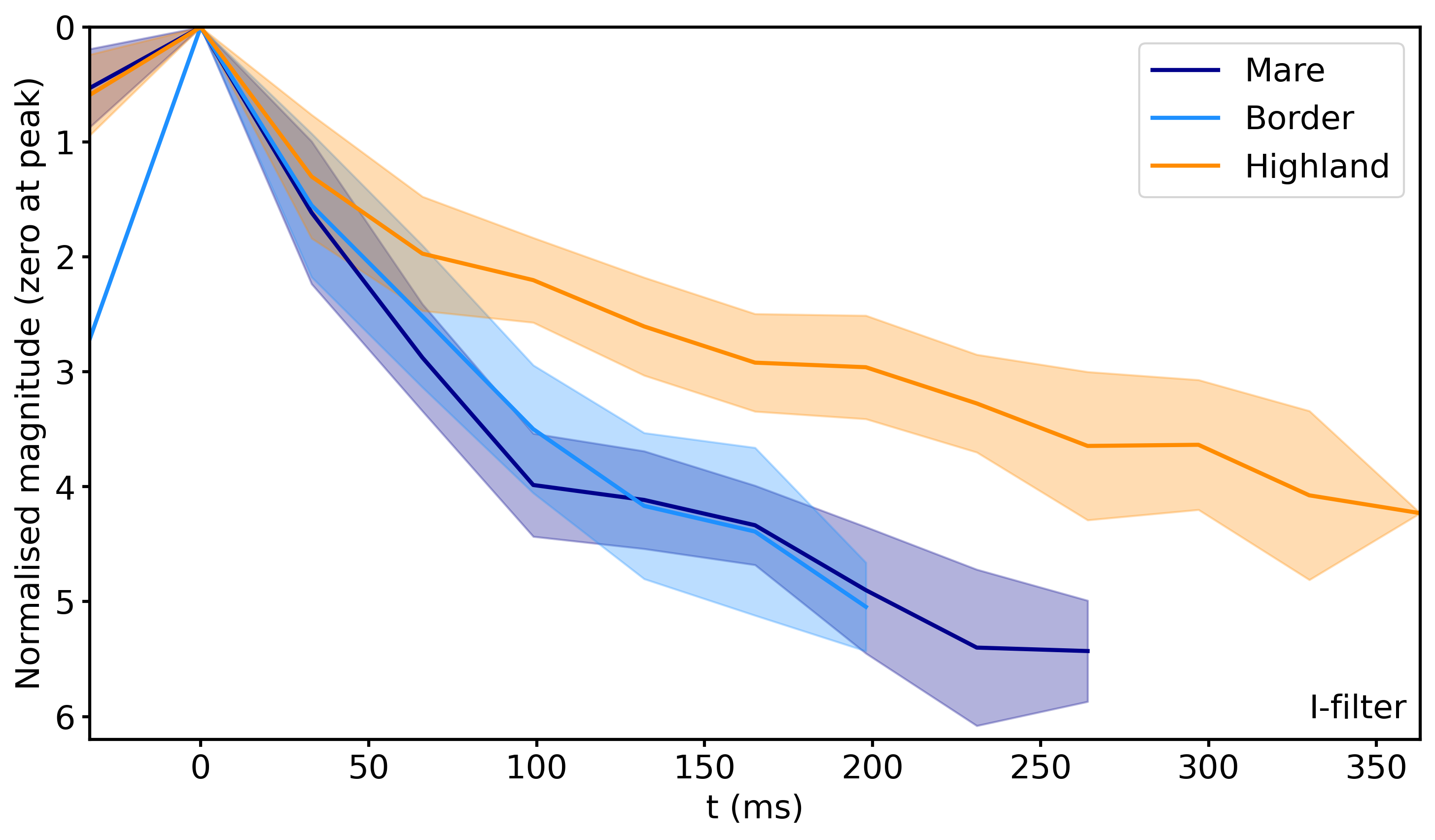}
    \includegraphics[width=0.44\textwidth]{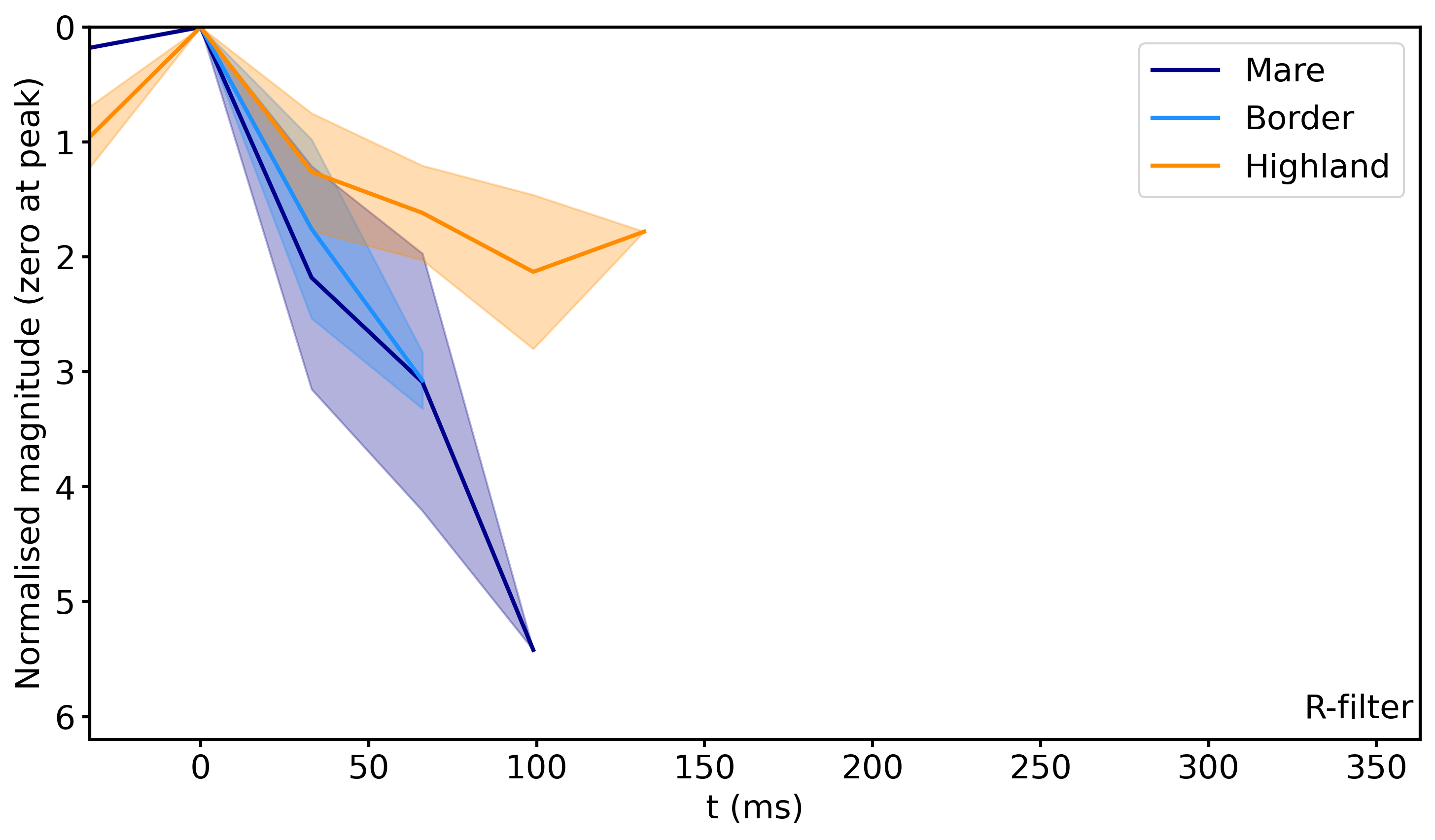}
      \caption{Average LIF light curves (including rms) for each terrain type in the $I$ filter (\textit{left}) and $R$ filter (\textit{right}).}
         \label{fig:LCs}
   \end{figure*}

\section{Ejecta cooling models}
The maximum luminosity of a LIF (i.e. the light curve peak) is primarily attributed to a vapour cloud or plasma radiation \citep{bouley2012, yanagisawa2025}. Laboratory experiments have demonstrated that the radiation emitted by the vapour cloud lasts less than 1~ms \citep{gores2022}. The post-peak emission is expected to originate from the ejecta material, extending the total flash duration from 10~ms up to 1~s for events within the 4--10 peak magnitude range \citep{bouley2012}.

In this work, we derived analytical expressions for the ejecta cooling models of \cite{yanagisawa2002} and \cite{yanagisawa2025}, adapting the mathematical formalism provided by \cite{bouley2012}. Notably, we departed from the use of the Stefan-Boltzmann law, as it represents the integrated flux over the entire spectrum; instead, we accounted for the band-specific response of photometric filters. Further details are provided in Appendix~\ref{app:CoolingModels}.

The absolute observed flux is described by 
\begin{equation}
F_\mathrm{\lambda}(t) = F_\mathrm{\lambda, 0} \cdot \Phi(t) ,
\label{eq:obsfluxsimple}
\end{equation}
where $F_{\lambda, 0}$ is a time-independent scaling factor that, when multiplied by $\Phi(t_\mathrm{ref}),$ yields the initial flux of the droplet model starting after the peak at the reference time ($t_\mathrm{ref}$), and $\Phi(t)$ is the light curve decay function defined as
\begin{equation}
   \Phi(t) = \left[e^{\frac{h c}{\lambda_\mathrm{eff} k_\mathrm{B} T_0}\left(1+ \frac{t-t_\mathrm{ref}}{\tau}\right)^{1/3}} - 1 \right]^{-1},
   \label{eq:ShapeFlux}
\end{equation}
where $h$ is Planck's constant, $c$ is the speed of light, $\lambda_\mathrm{eff}$ is the effective wavelength of the filter, $k_\mathrm{B}$ is the Boltzmann constant, $T_0$ is the initial temperature of the droplet model at time $t_\mathrm{ref}$, and $\tau$ is the characteristic cooling timescale, as defined in Appendix~\ref{app:CoolingModels}.

The normalised flux can be described by the following equation in the single-size model:
\begin{equation}
   F_\mathrm{norm}(t) = F_\mathrm{ref} \cdot \frac{\Phi(t)}{\Phi(t_\mathrm{ref})} + C_\mathrm{bg},
   \label{eq:NormFlux}
\end{equation}
where $F_\mathrm{ref} = \frac{F_\mathrm{\lambda}(t_\mathrm{ref})}{F_\mathrm{\text{peak}}}$ represents the fractional amplitude ($0 \le F_\mathrm{ref} \le 1$) of the data at the onset of the molten droplet phase. The $C_\mathrm{bg}$ is a mathematical baseline offset intended to account for non-zero instrumental noise. The shape of this normalised thermal tail depends entirely on $F_\mathrm{ref}$, $T_\mathrm{0}$, and $\tau$.

\cite{yanagisawa2025} suggested that two distinct particle populations are ejected during the excavation phase, both of which contribute to the LIF radiation. Initially, fine molten droplets form at temperatures near the vaporisation point of lunar minerals, which is followed by the ejection of coarse particles at lower initial temperatures. Due to their larger volume-to-surface-area ratio, these coarse particles cool much more slowly, acting as the primary source of the sustained afterglow. Following the same mathematical derivation as for the single-size model, Eq. (\ref{eq:NormFlux}) is expanded to account for both the fine ($\mathrm{f}$) and coarse ($\mathrm{c}$) populations:
\begin{equation}
   F_\mathrm{norm}(t) = F_\mathrm{ref} \cdot \frac{\Phi_\mathrm{f}(t) + A_\mathrm{R} \Phi_\mathrm{c}(t)}{\Phi_\mathrm{f}(t_\mathrm{ref}) + A_\mathrm{R} \Phi_\mathrm{c}(t_\mathrm{ref})} + C_\mathrm{bg},
   \label{eq:NormFluxDouble}
\end{equation}
where $A_\mathrm{R}$ is the cross-sectional area ratio between the coarse and fine particles.

In this work, we did not consider explicit regolith properties as fixed inputs. Instead, we employed a Markov chain Monte Carlo (MCMC) approach applied exclusively to the average $I$-filter LIF light curve (due to its extended duration compared to the $R$ filter) to explore the range of physical parameters in the analytical expressions (Eqs.~\ref{eq:NormFlux} and \ref{eq:NormFluxDouble}) that could explain the observational data (see Appendix~\ref{app:MCMC} for details).

\section{Results and discussion}

We plotted the average normalised light curves grouped by terrain type and find that they exhibit distinct morphologies (Fig.~\ref{fig:LCs}). A dichotomy between mare and highland LIFs becomes apparent approximately 50~ms after the peak. Highland LIFs decay more slowly than mare and `border' flashes, which show a similar behaviour. The dichotomy also persists when considering only validated LIFs, with the root mean square (rms) values being consistent with that of the full sample. Furthermore, this trend remains unchanged when comparing LIFs of the eastern and western hemispheres, or when isolating events that occur close to the lunar limb ($|\text{Long.}| > 65^\circ$).

While the general distribution of apparent peak magnitudes is similar across all groups (Fig.~\ref{fig:PeakMagDrop}, left), their initial magnitude decay differs significantly; mare and `border' LIFs exhibit an initial drop approximately twice as large as that observed for highland events (Fig.~\ref{fig:PeakMagDrop}, right). Furthermore, the long LIFs exhibit peaks that are $\sim$1.2~mag brighter on the mare and `border' regions than on the highlands (Fig.~\ref{fig:PeakMagDuration}). Consequently, a distinct magnitude offset occurs during the decay phase of the normalised light curves (Fig.~\ref{fig:LCs}), further widening the observed dichotomy. 

The extended duration and slower decay of highland LIFs can result in a luminous energy that is up to 30\% higher compared to a mare LIF of the same peak magnitude. If the impactor's kinetic energy is directly linked to the peak magnitude, then luminous efficiency will also be terrain-dependent, rather than solely a function of impactor velocity as suggested by \cite{swift2011} and \cite{fuse2020}. However, the divergent observed light curves suggest a difference in the initial energy partitioning. Mare LIFs, which feature a rapid magnitude drop, can lose a significant fraction of energy to the crushing and compaction of the denser basaltic target. Conversely, a highland LIF of equivalent energy can produce a fainter peak but more persistent emission because the impact energy is absorbed by the porous highland regolith, facilitating a more gradual thermal release.
\begin{figure}[ht!]
\centering
\includegraphics[width=\hsize]{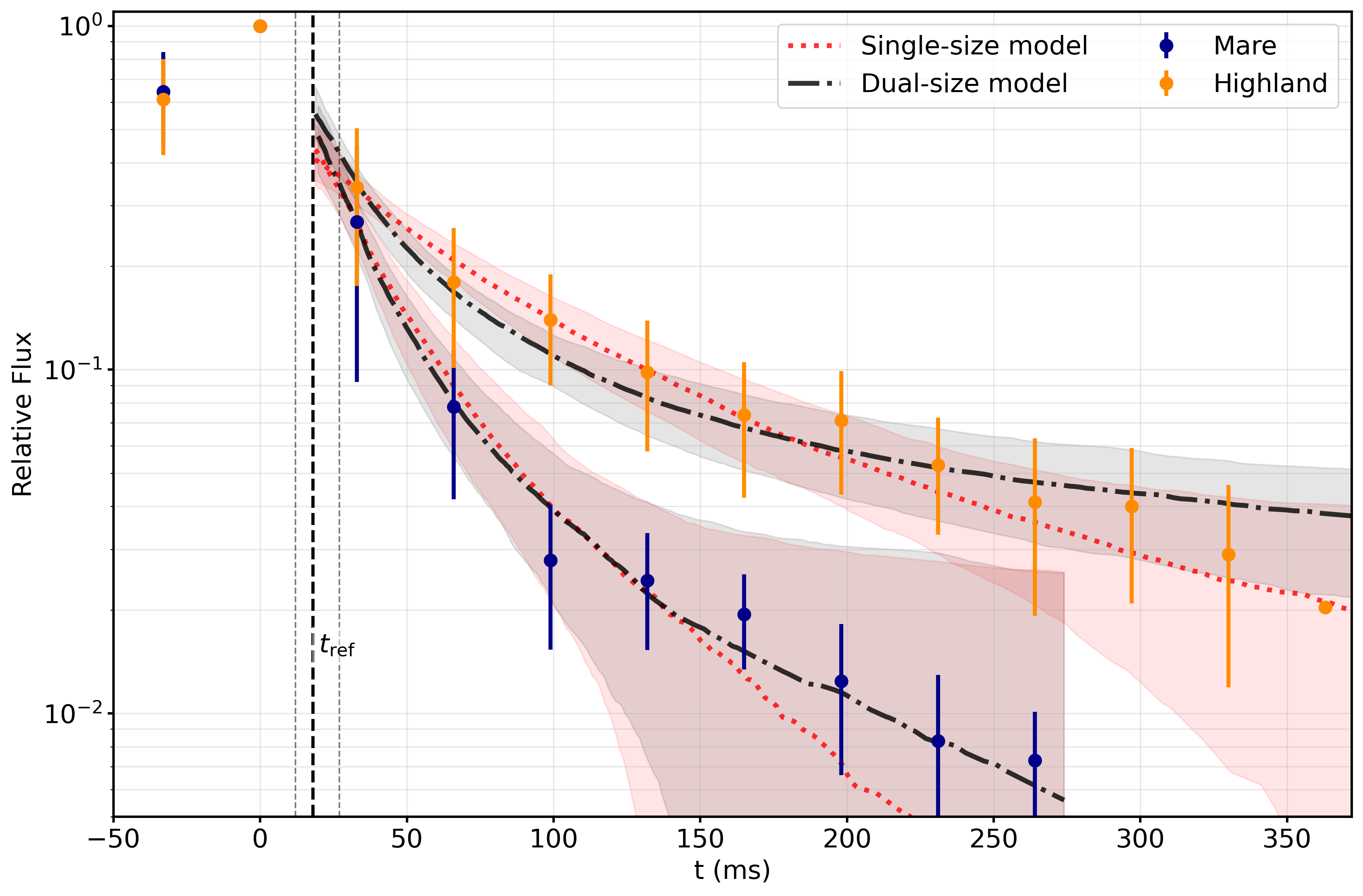}
\caption{Best-fit MCMC ejecta cooling models to the average (mare and highland) light curves (see Appendix~\ref{app:MCMC} for details).}
\label{fig:LIFmodels}
\end{figure}

The MCMC fitting of the single-size and dual-size thermal models to the NELIOTA data (Fig.~\ref{fig:LIFmodels}) uses the analytical expressions from Eqs.~(\ref{eq:NormFlux})~and~(\ref{eq:NormFluxDouble}), respectively. The final optimised parameters are summarised in Table~\ref{tab:MCMCresults}. When the single-size model is applied, the solutions for the mare and highland LIFs yield a comparable reference temperature of $\sim 3000$~K. However, they exhibit distinct cooling timescales: $\tau = 38_{-18}^{+32}$~ms for the maria and notably longer ($\tau = 110_{-50}^{+80}$~ms) for the highlands. The dual-size model predicts a stronger afterglow. For the fine droplet component, the two terrains exhibit similar reference temperatures in the range $3600 - 3700$~K, yet their cooling timescales differ slightly: $\tau_\mathrm{f} = 25_{-11}^{+14}$~ms for the maria compared to $\tau_\mathrm{f} = 36_{-13}^{+10}$~ms for the highlands. In contrast, the coarse particle component demonstrates disparate reference temperatures --- $T_\mathrm{c} = 2200_{-500}^{+600}$~K for the maria and $T_\mathrm{c} = 2700_{-600}^{+500}$~K for the highlands --- but its extended cooling timescale is the same for the two terrain types: $\tau_\mathrm{c} = 0.5_{-0.3}^{+0.3}$~s. Notably, the reference temperatures for both models are lower than $3800$~K, which is the estimated vaporisation temperature of the lunar regolith reported in previous studies \cite[][and references therein]{cintala1992}.

\section{Conclusions}
We analysed a sample of 124 multi-frame sporadic LIFs detected by NELIOTA and identified a dichotomy in their light curve behaviour as a function of terrain type. LIFs occurring on the highlands exhibit shallower and longer decays than those on the maria. `Border' LIFs display faster, steeper decay profiles that are similar to those of mare LIFs. The light curves were fitted using the newly derived analytical expressions based on the models of \cite{yanagisawa2002} and \cite{yanagisawa2025}, which quantify the initial temperatures and characteristic timescales for each terrain type. These expressions can be used to estimate the total luminous energy from location and temperature evolution using only a single band.

While studies have so far focused on the influence of surface lithology on late-stage crater morphology, this work provides the first evidence that the lunar terrain also plays a significant role in the initial energy partitioning during the earliest stages of the cratering process. The rapid magnitude drop observed in mare LIFs may be due to the denser basaltic target, where a significant fraction of the impact energy is consumed by the crushing and compaction of the consolidated rock. Conversely, the porous regolith of the highlands may facilitate a more gradual energy release, resulting in the persistent thermal tails observed in our sample. Whether these results indicate a fundamentally terrain-dependent luminous efficiency remains an open question. Analogous studies focusing on major meteoroid streams will be critical to providing an answer; in such cases, the impactor velocity and composition, parameters that also influence the initial temperature \citep{rackesh2026}, are more strictly constrained, allowing the influence of target lithology to be isolated.

The current study will serve as a highly valuable framework for the forthcoming Lunar Meteoroid Impacts Observer (LUMIO) CubeSat mission \citep{ferrari2025} of ESA, which will monitor the far side of the Moon for one year. Given that the far side is dominated by highlands, our results imply that LUMIO will detect more long LIFs. Future work will investigate correlations between surface parameters and observed fluxes, alongside temperature estimations derived via the $R$--$I$ colour index with the NELIOTA system \citep{bonanos2018, avdellidou2019, liakos2020}.

\begin{acknowledgements}
We are very grateful to the referee, M. Yanagisawa, for his careful reading of the paper. This work has made use of data from the NELIOTA project funded by the ESA. This study is based on observations made with the 1.2 m Kryoneri telescope, Corinthia, Greece, which is operated by the IAASARS, NOA. DA gratefully acknowledges financial support by the Horizon Europe Programme of the EU implemented by ESA under the NELIOTA-III programme, contract No. 4000148793/25/D/MRP. Views and opinion expressed are however those of the authors only and the European Commission cannot be held responsible for any use which may be made of the information contained therein.
\end{acknowledgements}
\bibliographystyle{bibtex/aa} 
\bibliography{Moon_bib}
\begin{appendix}
\nolinenumbers
\onecolumn
\section{Additional figures and tables}
\begin{table*}[!ht]
\centering
\caption{New LIFs observed by NELIOTA between August 2025 and February 2026.}
\label{tab:NewLIFs}
\small
\begin{tabular}{ccccccccc}
\hline \hline
ID & Valid. & Dur. (ms) & Date \& Time (UT) & $m_\mathrm{R,peak}$ (mag) & $m_\mathrm{I,peak}$ (mag) & Lat. ($^\circ$) & Long. ($^\circ$) & Source \\
\hline
296 & V   & 66  & 2025-08-17 00:28:47.619 & $10.2 \pm 0.3$  & $9.24 \pm 0.11$ & $-7.9 \pm 0.5$  & $54.6 \pm 0.5$ & SPO \\
297 & SC1 & 66  & 2025-08-17 00:36:58.774 &                 & $8.62 \pm 0.09$ & $46.0 \pm 0.5$  & $12.8 \pm 0.5$ & SPO \\
298 & SC1 & 66  & 2025-08-17 02:14:55.185 &                 & $9.33 \pm 0.10$ & $-13.7 \pm 0.5$ & $71.6 \pm 0.5$ & SPO \\
299 & SC1 & 66  & 2025-08-20 02:24:15.476 &                 & $9.20 \pm 0.08$ & $32.6 \pm 0.5$  & $-12.4 \pm 0.5$ & SPO \\
300 & SC2 & 33  & 2025-08-20 02:24:15.476 &                 & $10.02 \pm 0.12$& $51.9 \pm 0.5$  & $15.9 \pm 0.5$ & SPO \\
301 & V   & 132 & 2025-09-14 23:56:16.151 & $8.40 \pm 0.13$ & $7.63 \pm 0.07$ & $6 \pm 1$       & $72 \pm 1$     & SPO \\
302 & SC1 & 66  & 2025-09-15 01:03:26.148 &                 & $9.00 \pm 0.10$ & $-18 \pm 1$     & $62 \pm 1$     & SPO \\
303 & SC2 & 33  & 2025-09-16 02:27:13.313 &                 & $10.20 \pm 0.24$& $-37 \pm 1$     & $85 \pm 1$     & SPO \\
304 & SC2 & 33  & 2025-09-17 02:58:25.604 &                 & $9.52 \pm 0.09$ & $19 \pm 1$      & $68 \pm 1$     & SPO \\
305 & SC1 & 66  & 2025-10-28 16:31:32.388 &                 & $8.5 \pm 0.09$  & $18 \pm 1$      & $-83 \pm 1$    & SPO \\
306 & SC2 & 33  & 2025-10-28 16:46:51.240 &                 & $9.7 \pm 0.4$   & $6 \pm 1$       & $-29 \pm 1$    & ORI \\
307 & V   & 33  & 2025-10-28 18:32:01.270 & $7.58 \pm 0.11$ & $6.33 \pm 0.09$ & $-18 \pm 1$     & $-18 \pm 1$    & ORI \\
308 & SC1*& 33  & 2025-11-13 02:10:02.383 & $8.77 \pm 0.16$ & $8.70 \pm 0.13$ & $-27 \pm 1$     & $21 \pm 1$     & SPO \\
309 & V   & 66  & 2025-11-13 02:21:24.302 & $10.9 \pm 0.5$  & $8.81 \pm 0.09$ & $9 \pm 1$       & $70 \pm 1$     & SPO \\
310 & V   & 99  & 2025-11-13 03:10:06.033 & $10.04 \pm 0.24$& $8.47 \pm 0.07$ & $-5 \pm 1$      & $44 \pm 1$     & SPO \\
311 & V   & 33  & 2025-11-16 03:49:15.986 & $11.20 \pm 0.29$& $10.03 \pm 0.13$& $2.5 \pm 0.5$   & $12.5 \pm 0.5$ & SPO \\
312 & SC2 & 33  & 2025-11-16 03:55:52.917 &                 & $9.54 \pm 0.11$ & $9.4 \pm 0.5$   & $38.0 \pm 0.5$ & SPO \\
313 & V   & 66  & 2025-11-16 04:27:15.941 & $10.7 \pm 0.5$  & $9.49 \pm 0.24$ & $19.5 \pm 0.5$  & $50.8 \pm 0.5$ & SPO \\
314 & SC1 & 66  & 2025-11-24 16:41:24.558 &                 & $9.77 \pm 0.14$ & $22.9 \pm 0.5$  & $-28.8 \pm 0.5$ & SPO \\
315 & SC2 & 33  & 2025-12-13 04:26:10.455 &                 & $8.96 \pm 0.09$ & $14 \pm 1$      & $37 \pm 1$     & GEM \\
316 & SC2 & 33  & 2025-12-14 02:25:35.027 &                 & $10.31 \pm 0.17$& $13 \pm 1$      & $82 \pm 1$     & GEM \\
317 & V   & 99  & 2025-12-14 02:32:34.360 & $9.59 \pm 0.19$ & $8.08 \pm 0.09$ & $-14 \pm 1$     & $42 \pm 1$     & GEM \\
318 & V   & 66  & 2025-12-14 02:45:18.590 & $9.21 \pm 0.14$ & $8.15 \pm 0.08$ & $-24 \pm 1$     & $63 \pm 1$     & GEM \\
319 & V   & 33  & 2025-12-14 02:52:05.254 & $10.13 \pm 0.29$& $8.88 \pm 0.09$ & $12 \pm 1$      & $77 \pm 1$     & GEM \\
320 & SC2 & 33  & 2025-12-15 03:42:53.126 &                 & $9.94 \pm 0.16$ & $3.5 \pm 0.5$   & $20.8 \pm 0.5$ & GEM \\
321 & V   & 99  & 2025-12-15 03:51:33.930 & $10.7 \pm 0.3$  & $8.8 \pm 0.09$  & $-22.0 \pm 0.5$ & $12.5 \pm 0.5$ & GEM \\
322 & SC2 & 33  & 2025-12-15 04:16:54.026 &                 & $10.05 \pm 0.12$& $16.2 \pm 0.5$  & $68.4 \pm 0.5$ & GEM \\
323 & V   & 99  & 2025-12-15 04:33:42.042 & $9.24 \pm 0.10$ & $7.99 \pm 0.06$ & $19.1 \pm 0.5$  & $58.2 \pm 0.5$ & GEM \\
324 & SC2 & 33  & 2025-12-15 04:47:50.551 &                 & $10.16 \pm 0.13$& $-3.8 \pm 0.5$  & $65.9 \pm 0.5$ & GEM \\
325 & V   & 66  & 2026-01-13 04:09:55.147 & $9.16 \pm 0.11$ & $8.42 \pm 0.08$ & $-42.8 \pm 0.5$ & $48.4 \pm 0.5$ & SPO \\
326 & V   & 66  & 2026-02-21 17:31:42.570 & $9.03 \pm 0.13$ & $8.31 \pm 0.07$ & $-34 \pm 1$     & $-1 \pm 1$     & SPO \\
327 & SC2 & 33  & 2026-02-23 16:59:19.400 &                 & $7.76 \pm 0.08$ & $-12 \pm 1$     & $-11 \pm 1$    & SPO \\
328 & V   & 99  & 2026-02-23 17:58:19.746 & $9.6 \pm 0.8$   & $7.21 \pm 0.05$ & $15 \pm 1$      & $-18 \pm 1$    & SPO \\
329 & SC1 & 66  & 2026-02-23 19:52:19.732 &                 & $8.76 \pm 0.11$ & $-20 \pm 1$     & $-29 \pm 1$    & SPO \\
\hline
\end{tabular}
\tablefoot{The ID numbers continue from the previous NELIOTA study by \cite{liakos2024}. V = Validated LIF; SC1/2 = Suspected LIF of Class 1/2 \citep[see][for details]{liakos2024}; (*): denotes an abnormal $R-I$ index. The association of LIFs with active meteoroid streams was determined using the methods of \cite{madiedo2015b, madiedo2015a} and \cite{ortiz2015}.}
\end{table*}
\begin{figure*}[ht!]
   \centering
    \includegraphics[width=0.48\textwidth]{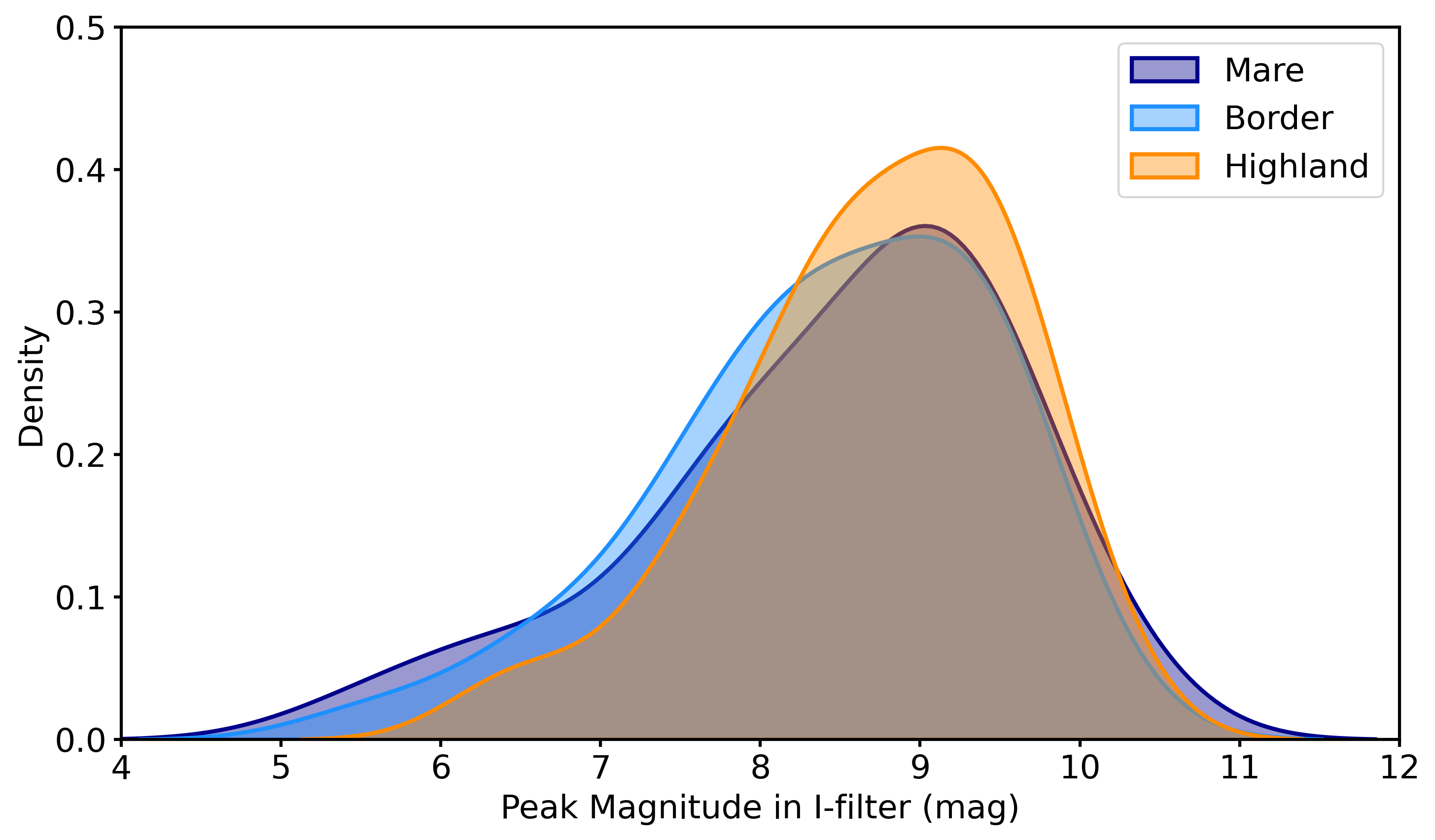}
    \includegraphics[width=0.48\textwidth]{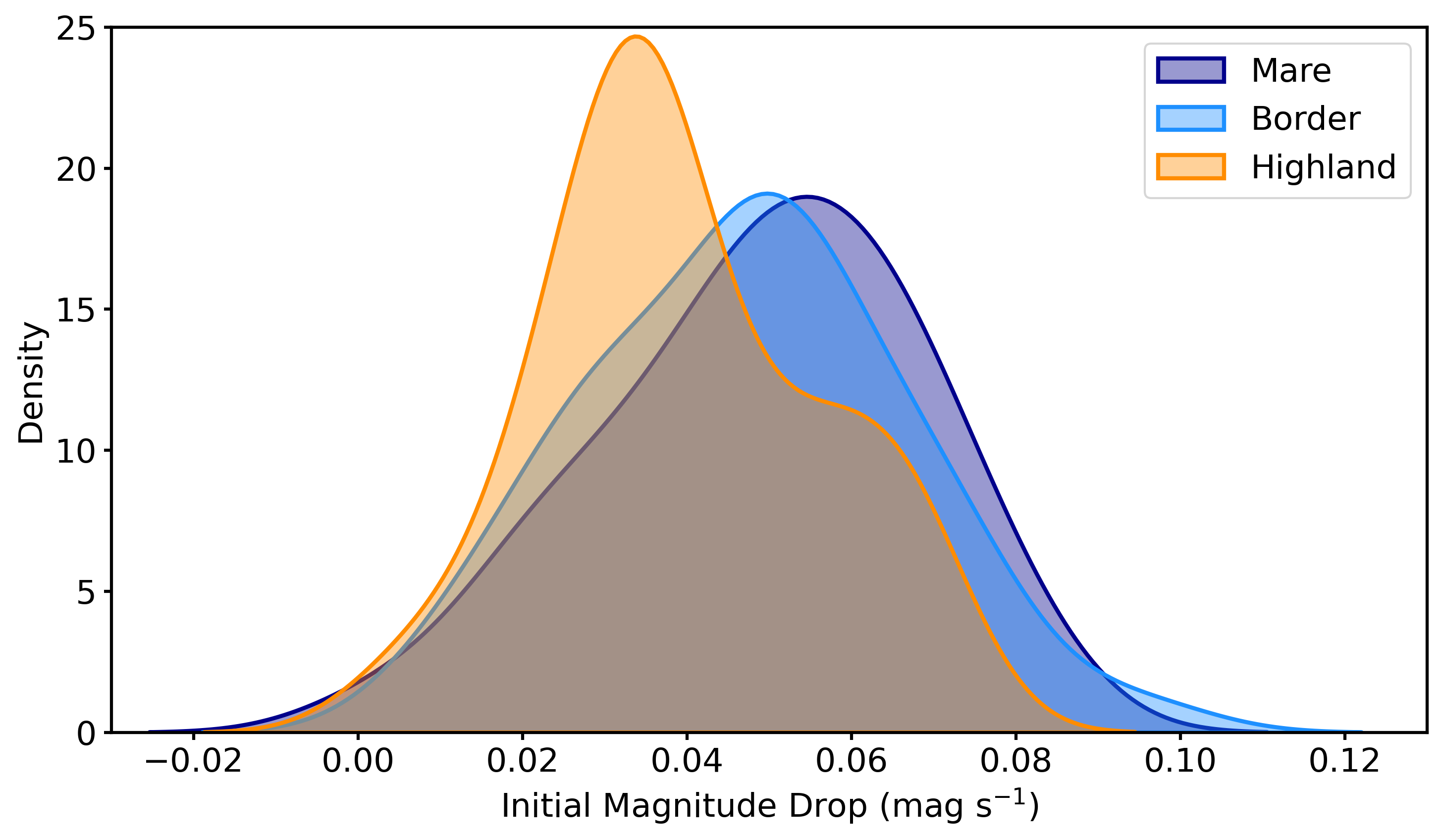}
      \caption{Kernel density estimation (KDE) of the peak magnitude in the I filter (\textit{left}) and of the initial magnitude drop (\textit{right}). For the KDE, a Gaussian kernel was applied using SciPy \citep{virtanen2020}, with the bandwidth determined according to Scott's rule \citep{scott2015}.}
         \label{fig:PeakMagDrop}
\end{figure*}

\begin{figure}[ht!]
\centering
\includegraphics[width=0.75\textwidth]{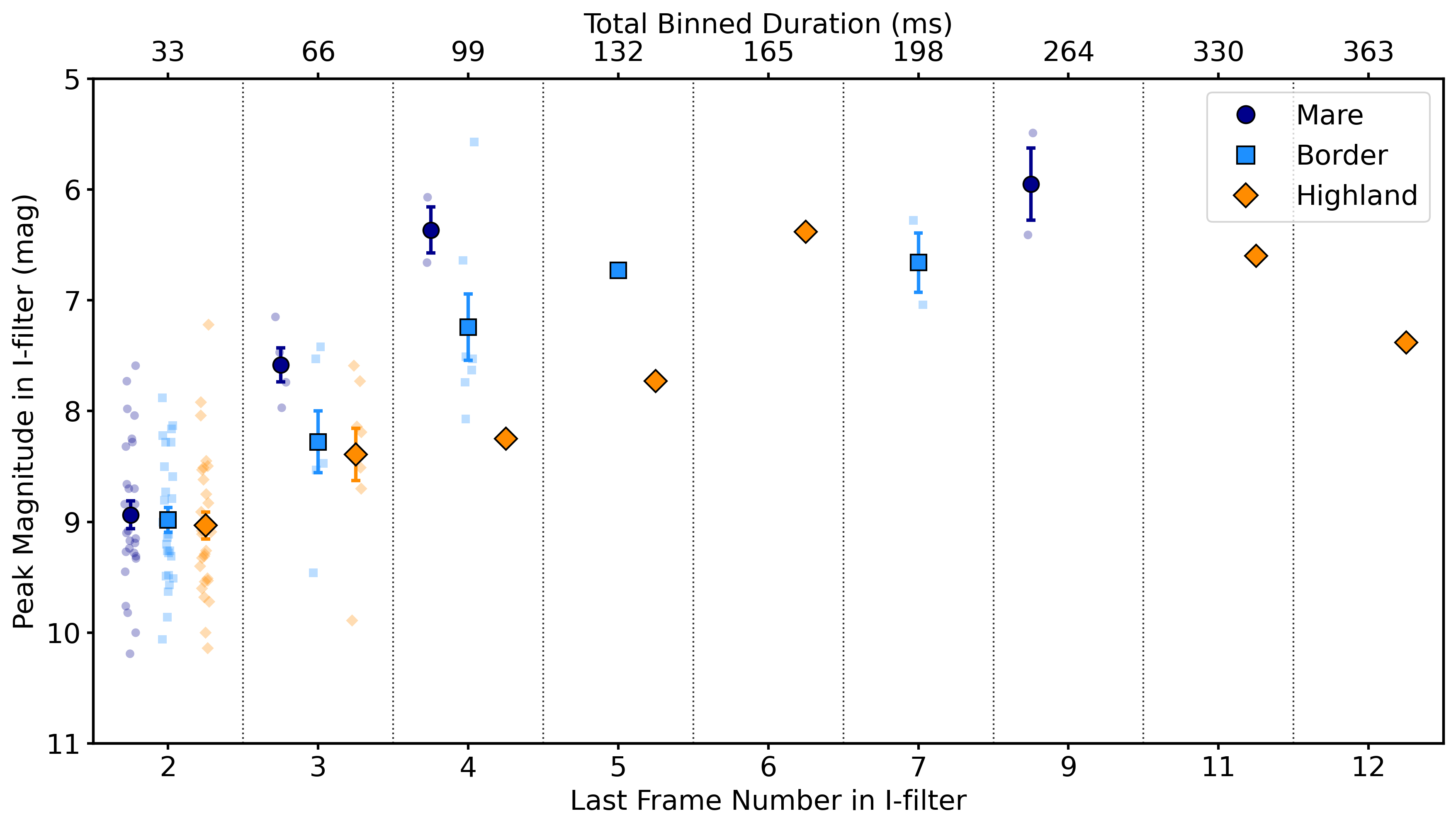}
\caption{Peak magnitude of the multi-frame LIFs in the $I$ filter as a function of their duration (in bins of 33~ms) and the last frame in which they were detected. Different colours correspond to different terrain types. Individual measurements are shown with faded markers, while bold markers represent their average within each bin. Frame 2 corresponds to the first measurement following the peak, whereas the undisplayed Frame 1 represents the measurement captured at the peak.}
\label{fig:PeakMagDuration}
\end{figure}
   
\begin{table*}[!ht]
\centering
\caption{Photometry of multi-frame sporadic LIFs observed by NELIOTA between August 2025 and February 2026.}
\label{tab:MultiframeLIFs}
\small
\begin{tabular}{cccc|cccc}
\hline \hline
ID & Frame & $m_\mathrm{R}$ (mag) & $m_\mathrm{I}$ (mag) & ID & Frame & $m_\mathrm{R}$ (mag) & $m_\mathrm{I}$ (mag) \\
\hline
296 & i   & $10.2 \pm 0.3$   & $9.24 \pm 0.1$   & 310 & i   & $10.04 \pm 0.24$ & $8.47 \pm 0.07$ \\
    & ii  &                  & $10.38 \pm 0.25$ &     & ii  &                  & $8.75 \pm 0.07$ \\
297 & i   &                  & $8.62 \pm 0.09$  &     & iii &                  & $10.3 \pm 0.4$ \\
    & ii  &                  & $9.72 \pm 0.22$  & 313 & i   & $10.7 \pm 0.5$   & $9.49 \pm 0.24$ \\
298 & i   &                  & $9.33 \pm 0.10$  &     & ii  &                  & $11.1 \pm 0.4$ \\
    & ii  &                  & $10.63 \pm 0.28$ & 314 & i   &                  & $9.77 \pm 0.14$ \\
299 & i   &                  & $9.20 \pm 0.08$  &     & ii  &                  & $11.1 \pm 0.4$ \\
    & ii  &                  & $9.87 \pm 0.12$  & 325 & i   & $9.16 \pm 0.11$  & $8.42 \pm 0.08$ \\
301 & i   &                  & $10.34 \pm 0.26$ &     & ii  &                  & $10.8 \pm 0.4$ \\
    & ii  & $8.40 \pm 0.12$  & $7.63 \pm 0.07$  & 326 & i   & $9.03 \pm 0.13$  & $8.31 \pm 0.07$ \\
    & iii & $9.93 \pm 0.26$  & $9.32 \pm 0.11$  &     & ii  &                  & $10.2 \pm 0.3$ \\
    & iv  &                  & $10.16 \pm 0.18$ & 328 & i   & $9.6 \pm 0.8$    & $7.21 \pm 0.05$ \\
    & v   &                  & $11.40 \pm 0.53$ &     & ii  &                  & $7.93 \pm 0.06$ \\
302 & i   &                  & $9.00 \pm 0.10$  &     & iii &                  & $10.0 \pm 0.4$ \\
    & ii  &                  & $10.8 \pm 0.4$   & 329 & i   &                  & $8.76 \pm 0.11$ \\
305 & i   &                  & $8.50 \pm 0.09$  &     & ii  &                  & $9.66 \pm 0.25$ \\
    & ii  &                  & $10.29 \pm 0.29$ &     &     &                  & \\
309 & i   & $10.9 \pm 0.5$   & $8.81 \pm 0.09$  &     &     &                  & \\
    & ii  &                  & $12 \pm 1$       &     &     &                  &\\
\hline
\end{tabular}
\end{table*}

\begin{table}[ht!]
\caption{Distribution of NELIOTA multi-frame sporadic LIFs vs terrain type.}                 
\label{tab:RegionStats}   
\centering            
\begin{tabular}{l c c c c}    
\hline\hline              
Terrain & Multi-frame & Pre-peak & Long\\        
\hline                     
   Mare     & 35 (31)   & 2 (2)  & 4 (4)  \\
   Border   & 42 (34)   & 1 (1)  & 10 (9) \\
   Highland & 42 (32)   & 4 (3)  & 5 (5) \\
   \hline
   Total    & 124 (97)   & 7 (6)  & 19 (18)\\
\hline                                 
\end{tabular}
\tablefoot{Numbers in parentheses correspond to the validated LIFs (those detected in both the $I$ and $R$ filters). `Pre-peak' refers to LIFs with measurements prior to the peak magnitude in the $I$ filter, while `long' refers to those with a duration longer than 100~ms. The multi-frame LIF data were adopted from Tables D.1 in \cite{liakos2020}, B.1 in \cite{liakos2024}, and \ref{tab:MultiframeLIFs} of this work.}
\end{table}

\begin{sidewaystable*}[!ht]
\centering
\caption{Terrain classification of multi-frame sporadic LIFs.}
\label{tab:LIFregions}
\small
\begin{tabular}{cccc|cccc|cccc|cccc}
\hline
\hline
ID & Valid. & Terrain & Mare  & ID & Valid. & Terrain & Mare  & ID & Valid. & Terrain & Mare  & ID & Valid. & Terrain & Mare  \\
& & type & Coverage (\%) & & & type &  Coverage (\%) & & & type & Coverage (\%) & &  & type & Coverage (\%) \\
\hline
2  & Yes & Mare     & 99  & 93  & Yes & Highland & 0   & 170 & No  & Border   & 93  & 259 & Yes & Highland & 0   \\
6  & Yes & Mare     & 99  & 94  & Yes & Mare     & 100 & 171 & Yes & Mare     & 100 & 261 & Yes & Highland & 0   \\
9  & No  & Highland & 0   & 95  & Yes & Highland & 0   & 172 & Yes & Border   & 96  & 264 & Yes & Mare     & 98  \\
12 & Yes & Mare     & 100 & 96  & Yes & Border   & 16  & 173 & No  & Border   & 10  & 266 & Yes & Highland & 0   \\
13 & Yes & Border   & 44  & 97  & Yes & Mare     & 100 & 177 & Yes & Mare     & 100 & 267 & Yes & Highland & 0   \\
14 & Yes & Mare     & 100 & 100 & Yes & Highland & 0   & 180 & Yes & Mare     & 99  & 269 & Yes & Highland & 0   \\
20 & Yes & Highland & 0   & 107 & Yes & Highland & 0   & 182 & Yes & Mare     & 100 & 270 & No  & Mare     & 100 \\
21 & Yes & Border   & 82  & 108 & Yes & Mare     & 100 & 184 & Yes & Highland & 0   & 272 & Yes & Border   & 77  \\
22 & Yes & Mare     & 100 & 118 & Yes & Highland & 0   & 185 & Yes & Mare     & 100 & 279 & Yes & Border   & 88  \\
23 & Yes & Border   & 61  & 124 & Yes & Highland & 0   & 186 & Yes & Highland & 0   & 280 & Yes & Border   & 81  \\
24 & Yes & Highland & 0   & 125 & Yes & Border   & 90  & 187 & No  & Highland & 0   & 281 & Yes & Mare     & 100 \\
26 & Yes & Border   & 10  & 126 & Yes & Border   & 47  & 188 & Yes & Mare     & 100 & 282 & Yes & Border   & 81  \\
27 & Yes & Border   & 57  & 127 & Yes & Border   & 36  & 191 & Yes & Border   & 6   & 283 & Yes & Border   & 22  \\
28 & Yes & Highland & 0   & 129 & No  & Border   & 62  & 195 & Yes & Border   & 94  & 284 & Yes & Border   & 97  \\
29 & Yes & Highland & 0   & 130 & Yes & Highland & 0   & 196 & No  & Mare     & 100 & 287 & Yes & Border   & 36  \\
30 & Yes & Border   & 75  & 133 & No  & Border   & 73  & 198 & No  & Highland & 0   & 289 & Yes & Border   & 46  \\
32 & Yes & Highland & 0   & 136 & Yes & Highland & 0   & 199 & Yes & Mare     & 100 & 291 & No  & Mare     & 100 \\
33 & Yes & Border   & 60  & 138 & No  & Border   & 71  & 202 & Yes & Highland & 0   & 296 & Yes & Mare     & 100 \\
35 & No  & Highland & 0   & 142 & Yes & Mare     & 100 & 204 & Yes & Border   & 10  & 297 & No  & Highland & 0   \\
47 & Yes & Border   & 20  & 144 & Yes & Mare     & 100 & 205 & Yes & Highland & 0   & 298 & No  & Highland & 0   \\
58 & Yes & Highland & 0   & 145 & Yes & Highland & 0   & 206 & Yes & Border   & 53  & 301 & Yes & Border   & 42  \\
59 & Yes & Highland & 0   & 148 & Yes & Border   & 70  & 217 & No  & Highland & 0   & 302 & No  & Border   & 44  \\
61 & Yes & Mare     & 100 & 150 & No  & Highland & 0   & 219 & Yes & Mare     & 100 & 305 & No  & Highland & 0   \\
62 & Yes & Border   & 49  & 151 & Yes & Mare     & 100 & 223 & Yes & Border   & 62  & 309 & Yes & Border   & 28  \\
73 & Yes & Highland & 0   & 152 & Yes & Mare     & 100 & 224 & Yes & Highland & 0   & 310 & No  & Border   & 85  \\
74 & Yes & Highland & 0   & 153 & Yes & Border   & 52  & 226 & Yes & Border   & 88  & 313 & Yes & Border   & 7   \\
76 & Yes & Border   & 7   & 160 & Yes & Mare     & 100 & 227 & No  & Border   & 39  & 314 & No   & Border   & 11  \\
78 & Yes & Mare     & 100 & 162 & Yes & Mare     & 100 & 235 & Yes & Highland & 0   & 325 & Yes   & Highland & 0   \\
80 & No  & Mare     & 100 & 163 & Yes & Highland & 0   & 252 & Yes & Highland & 0   & 326 & Yes   & Highland & 0   \\
85 & Yes & Mare     & 100 & 168 & Yes & Mare     & 100 & 253 & No  & Highland & 0   & 328 & Yes   & Border   & 91  \\
90 & Yes & Highland & 0   & 169 & Yes & Mare     & 100 & 257 & Yes & Mare     & 100 & 329 & Yes   & Border   & -     \\
\hline
\end{tabular}
\tablefoot{The LIFs with IDs from 2 to 108 are reported in \cite{liakos2020} and from 118 to 291 in \cite{liakos2024}. Mare coverage represents the percentage of the LIF area (defined by its localisation position and uncertainty) that overlaps with the mare region as defined by \cite{nelson2014}.}
\end{sidewaystable*}

\FloatBarrier
\twocolumn

\onecolumn
\section{LIF cooling models} \label{app:CoolingModels}
Laboratory experiments have shown that impacts are accompanied by the initial stages of impact crater formation; the contact and compression stage, and the excavation stage \citep{gores2022}. During the contact and compression stage, a vapour cloud or even plasma is produced, emitting radiation with a duration of less than 1~ms, as predicted by physical and numerical models \citep[e.g.][]{nemtchinov1998, artemeva2001, davis2009}. Various laboratory experiments have demonstrated this short timescale on lunar-analogue materials \citep[e.g.][]{ernst2003, spray2019, ernst2011, tandy2020, gores2022}. The emitted radiation during the second stage originates from the ejecta material, which cools down, causing a flash duration between 10~ms and $\approx 1$~s \citep{bouley2012}. \cite{yanagisawa2002} and \cite{yanagisawa2025} have developed the only LIF cooling models, also known as ejecta cooling models. In this work, we adapt these models to our data by incorporating their wavelength dependence, using the mathematical formalism of \cite{bouley2012} and \cite{song2025}.

\subsection{The single-size molten droplet model}

\cite{yanagisawa2002} assumed that as the vapour plume cools rapidly, small spherical droplets of a uniform size are formed and radiate as blackbodies. In a more general approach, we can model these droplets as greybodies. The thermal evolution of a spherical droplet can be described by the heat conduction equation
\begin{equation}
   \frac{\partial T}{\partial t} = \frac{k}{c_\mathrm{p} \rho} \frac{1}{r^2} \frac{\partial}{\partial r}\left(r^2 \frac{\partial T}{\partial r}\right),
   \label{eq:heatcond}
\end{equation}
with the boundary condition at the surface temperature $T_\mathrm{s}$:
$k \frac{\partial T}{\partial r} = - \epsilon \sigma T_\mathrm{s}^4$, 
where $k$ is the thermal conductivity, $c_\mathrm{p}$ is the specific heat capacity, $\rho$ is the droplet density, $\epsilon$ is the emissivity ($0 \le \epsilon \le 1$), and $\sigma$ is the Stefan-Boltzmann constant. 

Assuming a uniform temperature distribution within a droplet of radius $R_\mathrm{d}$, integrating Eq. (\ref{eq:heatcond}) over the droplet volume yields
\begin{equation}
   \frac{4}{3} \pi R_\mathrm{d}^3 c_\mathrm{p} \rho \frac{dT}{dt} = - 4 \pi R_\mathrm{d}^2 \epsilon \sigma T^4 \quad \Rightarrow \quad \frac{dT}{dt} = - \frac{3 \epsilon \sigma}{R_\mathrm{d} c_\mathrm{p} \rho} T^4 ,
\end{equation}
The analytical solution to this differential equation is
\begin{equation}
   T(t) = \frac{T_0}{\left( 1+ \frac{9 \epsilon \sigma T_0^3}{R_\mathrm{d} c_\mathrm{p} \rho} (t-t_\mathrm{ref}) \right)^{1/3}} \quad \Rightarrow \quad T(t) = T_0 \left( 1+ \frac{t-t_\mathrm{ref}}{\tau} \right)^{-1/3},
   \label{eq:thermalcurve}
\end{equation}
where $T_0$ is the initial temperature of the droplet at the onset time ($t_\mathrm{ref} \ge 0$) of the liquid phase, which may not necessarily coincide with the time of the absolute flux peak ($t_\mathrm{peak}=0$). For mathematical convenience, we can define the characteristic cooling timescale as $\tau = \frac{R_\mathrm{d} c_\mathrm{p} \rho}{9 \epsilon \sigma T_0^3}$ as \cite{bouley2012}.

The radiative flux of a greybody is described by Planck's law multiplied by the emissivity:
\begin{equation}
   f(\lambda,T) = \epsilon(\lambda) \frac{2 \pi h c^2}{\lambda^5} \frac{1}{e^{\frac{h c}{\lambda k_\mathrm{B} T}} - 1},
\end{equation}
where $\lambda$ is the wavelength, $h$ is the Planck constant, $c$ is the speed of light and $k_\mathrm{B}$ is the Boltzmann constant. Assuming the emissivity is independent of wavelength ($\epsilon(\lambda)=\epsilon$) and the optical filter has a uniform response over a bandwidth $\Delta \lambda$ at an effective wavelength $\lambda_\mathrm{eff}$, the spectral radiance becomes
\begin{equation}
   f_\mathrm{\lambda}(T) \approx \epsilon \frac{2 \pi h c^2}{\lambda_\mathrm{eff}^5} \Delta \lambda \frac{1}{e^{\frac{h c}{\lambda_\mathrm{eff} k_\mathrm{B} T}} - 1} 
   \mathrel{\mathop{\Rightarrow}^{\ref{eq:thermalcurve}}}
   f_\mathrm{\lambda}(t) = \epsilon \frac{2 \pi h c^2}{\lambda_\mathrm{eff}^5} \Delta \lambda \left[ e^{\frac{h c}{\lambda_\mathrm{eff} k_\mathrm{B} T_0}\left(1+ \frac{t-t_\mathrm{ref}}{\tau}\right)^{1/3}} - 1 \right]^{-1}.
\end{equation}
The total luminosity is described by $L_\mathrm{\lambda}(t) = A_\mathrm{eff} f_\mathrm{\lambda}(t)$, where the effective emitting area ($A_\mathrm{eff}$) of the ejecta can be estimated as $A_\mathrm{eff} = \frac{3V_\mathrm{ej}}{R_\mathrm{d}}$, with $V_\mathrm{ej}$ representing the total ejecta melt volume, which is related to the mass and the impact velocity of the projectile \citep[see e.g.][and references therein]{song2025}.

Combining these relations, the absolute observed flux as a function of time is
\begin{equation}
   F_\mathrm{\lambda}(t) = \frac{L_\mathrm{\lambda}(t)}{f \pi D^2} \quad \Rightarrow \quad
   F_\mathrm{\lambda}(t) = \underbrace{\epsilon \frac{6 V_\mathrm{ej} h c^2}{f D^2 R_\mathrm{d} \lambda_\mathrm{eff}^5} \Delta \lambda}_{F_\mathrm{\lambda, 0}} \underbrace{\left[e^{\frac{h c}{\lambda_\mathrm{eff} k_\mathrm{B} T_0}\left(1+ \frac{t-t_\mathrm{ref}}{\tau}\right)^{1/3}} - 1 \right]^{-1}}_{\Phi(t)} 
   \quad \Rightarrow \quad F_\mathrm{\lambda}(t) = F_\mathrm{\lambda, 0} \cdot \Phi(t),
   \label{eq:ObservedFlux}
\end{equation}
where $f$ is a geometric factor ($2 \le f \le 4$) depending on the radiation anisotropy \citep[see e.g.][]{bouley2012}, and $D$ is the distance between the Earth and the lunar surface.

To fit this theoretical framework on our data, which are normalised such that the absolute maximum of the initial flash is unity ($F_\mathrm{peak} = 1$), we must isolate the thermodynamic cooling profile from the unknown physical parameters embedded in $F_\mathrm{\lambda, 0}$ (namely, volume, distance, and droplet radius).

By evaluating Eq.~(\ref{eq:ObservedFlux}) exactly at the reference time $t_0$, we obtain $F_\mathrm{\lambda}(t_\mathrm{ref}) = F_\mathrm{\lambda, 0} \cdot \Phi(t_\mathrm{ref})$. Taking the ratio of the time-dependent flux to this reference flux perfectly cancels the term $F_\mathrm{\lambda, 0}$, yielding the dimensionless cooling shape:
\begin{equation}
   \frac{F_\mathrm{\lambda}(t)}{F_\mathrm{\lambda}(t_\mathrm{ref})} = \frac{\Phi(t)}{\Phi(t_\mathrm{ref})}.
\end{equation}

To map this theoretical shape onto our normalised dataset, we scale it by an amplitude parameter $F_\mathrm{ref}$, yielding the final fitting function:
\begin{equation}
   F_\mathrm{norm}(t) = F_\mathrm{ref} \cdot \frac{\Phi(t)}{\Phi(t_\mathrm{ref})} + C_\mathrm{bg},
   \label{eq:NormFluxApp}
\end{equation}
where $F_\mathrm{ref} = \frac{F_\mathrm{\lambda}(t_\mathrm{ref})}{F_\mathrm{peak}}$ represents the fractional amplitude ($0 \le F_\mathrm{ref} \le 1$) of the data at the onset of the molten droplet phase. Additionally, $C_\mathrm{bg}$ is a bounded free parameter ($-0.05 \le C_\mathrm{bg} \le 0.05$) representing a mathematical baseline offset intended to account for residual, non-zero instrumental noise. The shape of this normalised thermal tail depends entirely on $F_\mathrm{ref}$, $T_0$, and $\tau$.

\subsection{The dual-size molten droplet model}
\cite{yanagisawa2025} suggested that two distinct particle populations are ejected during the excavation phase, both contributing to the LIF radiation. Initially, fine molten droplets form at temperatures near the vaporisation point of lunar minerals, followed by the ejection of coarse particles at lower initial temperatures. Due to their larger volume-to-surface-area ratio, these coarse particles cool much more slowly, acting as the primary source of the flux tail, the so-called `afterglow'.

Following the same mathematical derivation as the single-size model, Eq. (\ref{eq:NormFluxApp}) is expanded to account for both the fine ($\mathrm{f}$) and coarse ($\mathrm{c}$) populations:
\begin{equation}
   F_\mathrm{norm}(t) = F_\mathrm{ref} \cdot \frac{\Phi_\mathrm{f}(t) + A_\mathrm{R} \Phi_\mathrm{c}(t)}{\Phi_\mathrm{f}(t_\mathrm{ref}) + A_\mathrm{R} \Phi_\mathrm{c}(t_\mathrm{ref})} + C_\mathrm{bg},
   \label{eq:NormFluxDoubleApp}
\end{equation}
where $A_\mathrm{R}$ is the cross-sectional area ratio between the coarse and fine particles. This parameter is directly related to their total volume ratio via the expression $\frac{V_\mathrm{c}}{V_\mathrm{f}} = A_\mathrm{R} \frac{R_\mathrm{c}}{R_\mathrm{f}}$.
Physically, impact cratering kinematics dictate that the hotter fine droplets are ejected slightly earlier than the cooler coarse aggregates. However, both populations are assumed to start simultaneously at the observational reference time ($t_\mathrm{ref}$), which may lead to a slight overestimation of $T_\mathrm{c}$. This is negligible given our current resolution.

\section{Markov chain Monte Carlo model fitting}
\label{app:MCMC}
We applied a MCMC approach to the $I$-filter average LIF light curve, owing to its extended duration in comparison with the $R$ filter. In Eq.~(\ref{eq:ShapeFlux}), the effective wavelength of the $I$ Cousins filter was set to $\lambda_\mathrm{eff} = 798$~nm \citep[][]{bonanos2018}. The parameter space was explored using the \textsc{emcee} Python package \citep{foreman-mackey2013}. For the single-size model, the chains were evolved using 128 walkers running for 10,000 steps, with the first 5,000 steps discarded as a burn-in period to ensure convergence. To account for higher dimensionality and parameter degeneracy of the dual-size model, these computational parameters (walkers, total steps, and burn-in) were increased fourfold. Strict hierarchical physical priors (e.g. $T_\mathrm{f} > T_\mathrm{c} + 100$~K and $\tau_\mathrm{f} < \tau_\mathrm{c}$) were applied to the dual-size model to prevent population inversion. The final parameter estimates and their asymmetric $1\sigma$ credible intervals were derived from the median ($50^{\text{th}}$ percentile) and the $16^{\text{th}}$ and $84^{\text{th}}$ percentiles of the marginalised posterior distributions, respectively.

\begin{table*}[!hb]
\centering
\caption{MCMC results for the free parameters of each model.}
\label{tab:MCMCresults}
\renewcommand{\arraystretch}{1.5}
\begin{tabular}{lcccccccc}
\hline \hline
\multicolumn{9}{c}{Single-size model free parameters}  \\ \hline                                                              
Region   & $t_\mathrm{ref}$ (s)& $F_\mathrm{ref}$ & & \multicolumn{2}{c}{$T_\mathrm{ref}$ (K)}  & \multicolumn{2}{c}{$\tau$ (s)} & $C_\mathrm{bg}$   \\\hline
Mare     & $0.018_{-0.006}^{+0.007}$  & $0.51_{-0.12}^{+0.23}$   & & \multicolumn{2}{c}{$3000_{-700}^{+1000}$} & \multicolumn{2}{c}{$0.038_{-0.018}^{+0.032}$}  & $0.001_{-0.026}^{+0.024}$  \\
Border   & $0.018_{-0.006}^{+0.007}$  & $0.48_{-0.10}^{+0.21}$   & & \multicolumn{2}{c}{$3000_{-700}^{+900}$} & \multicolumn{2}{c}{$0.050_{-0.023}^{+0.039}$}  & $-0.004_{-0.030}^{+0.031}$ \\
Highland & $0.018_{-0.006}^{+0.008}$  & $0.42_{-0.05}^{+0.09}$   & & \multicolumn{2}{c}{$3100_{-800}^{+900}$} & \multicolumn{2}{c}{$0.11_{-0.05}^{+0.08}$}  & $0.008_{-0.035}^{+0.063}$  \\ 
\hline
\multicolumn{9}{c}{Dual-size model free   parameters}   \\ \hline
Region   & $t_\mathrm{ref}$ (s) & $F_\mathrm{ref}$ & $A_\mathrm{R}$  & $T_\mathrm{f}$ (K) & $T_\mathrm{c}$ (K)          & $\tau_\mathrm{f}$ (s)   & $\tau_\mathrm{c}$ (s)   & $C_\mathrm{bg}$   \\ \hline
Mare     & $0.020_{-0.006}^{+0.007}$   & $0.56_{-0.15}^{+0.22}$ & $0.4_{-0.3}^{+0.4}$ & $3600_{-700}^{+600}$   & $2200_{-500}^{+600}$  & $0.025_{-0.011}^{+0.014}$   & $0.5_{-0.3}^{+0.3}$   & $-0.003_{-0.026}^{+0.024}$ \\
Border   & $0.019_{-0.006}^{+0.007}$   & $0.55_{-0.14}^{+0.22}$ & $0.5_{-0.3}^{+0.4}$ & $3700_{-700}^{+600}$   & $2300_{-500}^{+600}$  & $0.030_{-0.013}^{+0.012}$   & $0.5_{-0.3}^{+0.3}$       & $-0.003_{-0.028}^{+0.028}$ \\
Highland & $0.018_{-0.006}^{+0.008}$   & $0.57_{-0.14}^{+0.19}$ & $0.6_{-0.3}^{+0.3}$ & $3700_{-700}^{+500}$   & $2700_{-600}^{+500}$  & $0.036_{-0.013}^{+0.010}$   & $0.5_{-0.3}^{+0.3}$   & $0.020_{-0.035}^{+0.020}$ \\
\hline
\end{tabular}
\end{table*}

\FloatBarrier
\twocolumn

\end{appendix}
\end{document}